\documentclass[12pt]{article}

\newcommand{\x}{arXiv:}
\newcommand{\m}{\mathrm}

\usepackage{graphicx}

\begin{document}
\thispagestyle{empty}
\begin{center}

\null \vskip-1truecm \vskip2truecm

{\Large{\bf \textsf{Holography of the Quark Matter Triple Point}}}

{\large{\bf \textsf{}}}

{\large{\bf \textsf{}}}

\vskip1truecm

{\large \textsf{Brett McInnes
}}

\vskip1truecm

\textsf{\\  National
  University of Singapore}

\textsf{email: matmcinn@nus.edu.sg}\\

\end{center}
\vskip1truecm \centerline{\textsf{ABSTRACT}} \baselineskip=15pt
\medskip

The quark matter phase diagram is believed to contain two distinguished points, lying on the boundary of the Quark-Gluon Plasma phase: a critical point and a triple point. In the holographic [``AdS/QCD"] approach, the region of relatively low chemical potentials around the phase transition near the critical point may be described using generalizations of the Hawking-Page transition. We propose that the \emph{other} QGP phase line, beginning at the triple point and rising towards the region of extremely high temperatures and chemical potentials, is described instead by a non-perturbative string effect discovered by Seiberg and Witten. Using an assumed position for the critical point, we are able to use this proposal to obtain a holographic lower bound on the temperature of the triple point. Combined with Shuryak's upper bound on this temperature, this leads to a rough estimate of the location of the triple point, at a temperature of around 70 MeV, and a chemical potential of about 1100 MeV.

\newpage

\addtocounter{section}{1}
\section* {\large{\textsf{1. Towards a Holographic Quark Matter Phase Diagram}}}

A key objective of theoretical and experimental investigations of QCD is to understand the \emph{quark matter phase diagram}\footnote{See \cite{kn:halasz}; also \cite{kn:alford} for an in-depth review, and \cite{kn:mohanty} for a more recent summary of what is known about the diagram.}. This displays the various phases of quark matter as the temperature T and the [baryonic] chemical potential $\mu$ are varied; a [much simplified] version is portrayed in Figure 1, which follows \cite{kn:alford}.
\begin{figure}[!h]
\centering
\includegraphics[width=1.1\textwidth]{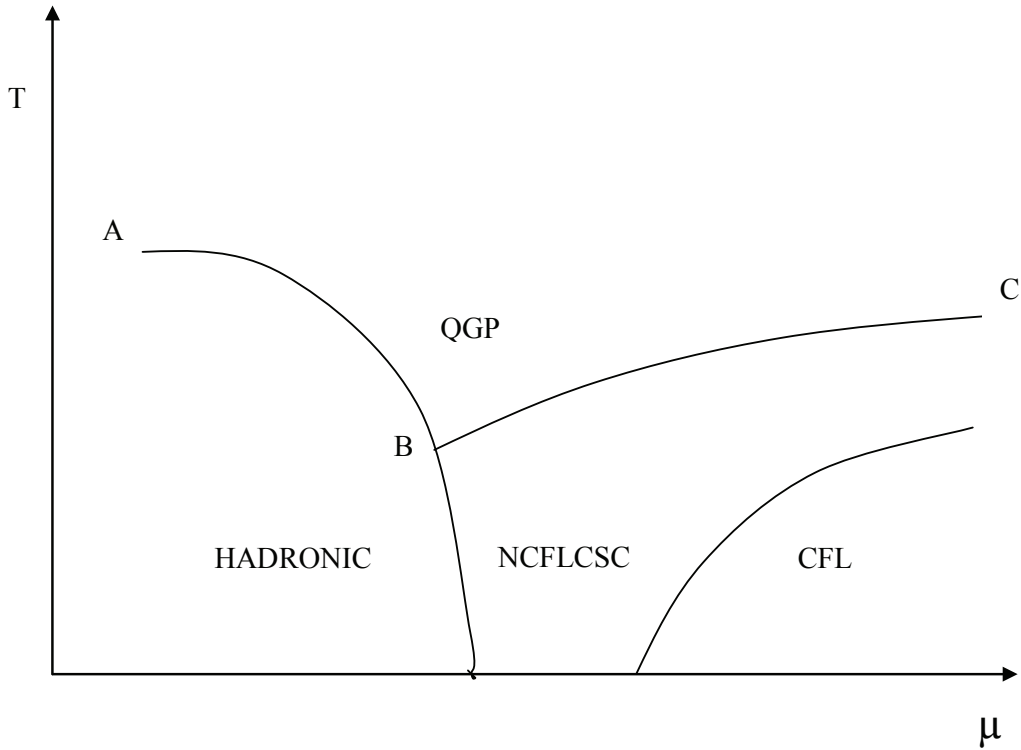}
\caption{Schematic Quark Matter Phase Diagram [after Alford et al. \cite{kn:alford}]; NCFLCSC = Non-CFL Colour SuperConductor.}
\end{figure}

Hadronic matter [which itself takes on various phases, not shown in Figure 1] occupies the region of relatively small T and small $\mu$. Remaining with small $\mu$ and increasing the temperature, one eventually reaches a region corresponding to one form of deconfined quark matter, known as the \emph{Quark-Gluon Plasma} or QGP. On the other hand, at low T but very high $\mu$, one finds a still more exotic form of matter called, in \cite{kn:alford}, the ``non-CFL colour superconducting" phase [or perhaps the recently proposed ``quarkyonic" phase \cite{kn:larry}]; beyond this lies the CFL or ``Colour-Flavour Locked" phase. The details will not concern us here; the key points for us are the following.

First, the QGP phase does not reach or come near to the T = 0 axis. This may seem obvious ---$\,$ the plasma cannot be [arbitrarily] cold ---$\,$ but this simple observation actually raises important theoretical problems, as has recently been stressed by Hartnoll [\cite{kn:hartnollreview}, footnote 14; see also \cite{kn:hartnollagain}]; we shall discuss this point in detail below. Second, at low values of $\mu$ one does \emph{not} find a sharp phase transition as the temperature decreases and hadronization begins to set in: instead there is a crossover \cite{kn:mohanty}, which may be narrow but which is not negligible. Correspondingly, there is a critical point at A in Figure 1. Thirdly, the presence of at least three phases leads to the existence of a \emph{triple point}, represented by B in Figure 1; on general grounds this is expected to have a lower temperature, and of course a higher chemical potential, than the critical point. This triple point is of considerable theoretical, and perhaps observational interest: one needs to fix its position in order to locate the two lines AB and BC which cross there\footnote{Of course, the phase line BC continues on indefinitely beyond the point C, which has been inserted merely for convenience.}, and furthermore this point corresponds to the \emph{lowest possible temperature} of the QGP. Determining this lowest possible temperature is an important goal. Fourthly and finally, we see that the line BC rises indefinitely, so that, at large values of $\mu$, there is a phase transition away from the QGP \emph{even at extremely high temperatures.}

Experiments with heavy ion collisions [for example, \cite{kn:phobos}\cite{kn:others}\cite{kn:wied}] explore the region near to the T axis in Figure 1; so does early-universe cosmology [since the matter/antimatter asymmetry, measured by the baryonic chemical potential, was small at very early times]; and so do most lattice calculations in QCD \cite{kn:fodor}. Also, this is the region to which the ``holographic" gauge-gravity correspondence was first applied \cite{kn:wittenads}, and where it continues to enjoy a remarkable degree of success \cite{kn:gubserreview}. On the other hand, the region near to the $\mu$ axis is the domain of relatively cold matter at extreme pressures; one hopes to probe this region by means of careful observations of neutron stars \cite{kn:alford2}, the cores of which may well consist of such forms of quark matter. Again, holographic methods have been applied successfully to part of this region: see for example \cite{kn:rozali} and references to it, and \cite{kn:chen} for the holography of the CFL phase.

The region where neither T nor $\mu$ is small, however, remains poorly understood, both theoretically and observationally. Indeed, it must be stressed that Figure 1 is, as indicated, ``schematic": there is a general consensus regarding the overall shape of the diagram, but far less agreement on the \emph{numerical} details [see for example \cite{kn:stephanov}]. In particular, the precise locations of the critical point A and the triple point B continue to be matters of considerable uncertainty, both theoretically and observationally; mapping out the phase diagram in the vicinity of these points has long been a major experimental project \cite{kn:raja}. Because it occurs at relatively low values of $\mu$, the critical point lies in a region which will probably be probed experimentally in the relatively near future ---$\,$ certainly, much more is known about it than about the triple point. [Some preliminary lattice calculations attempting to locate the latter have been done \cite{kn:triple}. Experimental probes of the region of the phase diagram in the direction of the triple point are currently in the initial stages of preparation \cite{kn:hohne}. For evidence of the existence of a triple point in connection with the proposed ``quarkyonic" phase, see \cite{kn:andronic}.]

One naturally hopes to bring holographic techniques to bear on the boundary of the QGP phase, defined by the lines AB and BC. As with all other theoretical approaches to quark matter, however, there are daunting obstacles to be overcome. For example, one does not know precisely how to represent a crossover holographically\footnote{This problem was discussed in \cite{kn:pufu}, but only the $\mu$ = 0 case is considered there.}. On the whole, however, recent progress makes it reasonable to hope that the theory of ``holographic superconductivity\footnote{See for example \cite{kn:horror}\cite{kn:horrorrob}\cite{kn:rocha}\cite{kn:kachru}\cite{kn:aprile}\cite{kn:pani}.}" will ultimately be able to account for the part of the phase diagram corresponding to temperatures and values of the chemical potential around those of the critical point A and its attached phase line AB.

As we shall discuss in the next section, however, there is also reason to think that the line BC in Figure 1 does \emph{not} correspond to any of the familiar holographic effects [defined by suitable generalizations of the Hawking-Page transition] which account for phase transitions at lower values of the chemical potential. In this work, we shall argue that there is a strictly stringy, non-perturbative effect which can account for BC. It does so by causing certain otherwise innocuous black holes to become unstable; we therefore begin with a discussion of the role of black hole instability.

\addtocounter{section}{1}
\section* {\large{\textsf{2. The Role of Black Hole Instability}}}
The part of the phase diagram which is best understood is the region corresponding to the QGP, the part lying above the lines AB and BC in Figure 1. The QGP is represented holographically by an asymptotically AdS Reissner-Nordstr$\ddot{\m{o}}$m thermal black hole, in which the electric charge corresponds holographically to the baryonic chemical potential \cite{kn:chamblin}. Black holes tend to have ``universal" behaviour, as for example their ``no hair" properties.
By focussing on the field theory phase which has an AdS Reissner-Nordstr$\ddot{\m{o}}$m dual, we can try to benefit from this universality. That is, it may be reasonable to expect that results based on the behaviour of black holes are broadly applicable even to field theories which differ from the specific ones that arise in the AdS/CFT correspondence; we thus hope to avoid
the perils \cite{kn:perils}\cite{kn:erlich} that beset all attempts to use the correspondence in this manner.

The natural way for a black hole description to fail is for the black hole itself to become \emph{unstable} as its temperature is lowered. That such instabilities can arise and have an important physical interpretation became clear with Witten's celebrated discussion \cite{kn:wittenads} of the Hawking-Page transition: sufficiently cold uncharged AdS black holes are indeed unstable to a transition to a different geometry, representing holographically the transition from a deconfined to a confined field theory. Witten's discussion extends straightforwardly to the charged case, but it immediately becomes apparent that this particular form of instability does \emph{not} explain what one sees in Figure 1. For the resulting phase diagram [for black holes with \emph{spherical} event horizons ---$\,$ see \cite{kn:clifford}, page 465] suggests that, at larger values of $\mu$, the plasma phase should extend down \emph{arbitrarily} close to the T = 0 axis. Apart from being intrinsically unreasonable, this means that there is no third phase and of course no triple point.

In fact we can turn this observation around, and assert that holography, combined with Figure 1, is already making an important prediction: the particular black holes that are dual to these field theories must \emph{always} [that is, at all values of $\mu$] become unstable as they are cooled, before they can reach zero temperature. This resolves, at least in the context of this application, the well-known puzzle of the apparent conflict between the physics of extremal black holes and the third law of thermodynamics. Of course, it remains to understand exactly \emph{why} these black holes become unstable when they are cooled, even at large values of the chemical potential: as we have seen, the Hawking-Page transition apparently does not explain this.

\emph{A key difficulty here is that the line BC in Figure 1 actually rises indefinitely.} In the holographic picture, this means that certain highly charged, extremely hot black holes [corresponding to points just below BC, near and beyond C in the figure] must be unstable. This presents a challenge, because most of the known forms of AdS black hole instability follow the pattern of the Hawking-Page transition: they only set in at low temperatures, particularly when the electric charge is large. This is in fact what one expects, since asymptotically AdS black holes with large electric charges have positive specific heats [see \cite{kn:clifford}, page 461]. Very hot black holes of this kind are very massive and have large entropies, and can reach thermal equilibrium with their own Hawking radiation: they are very different from the familiar asymptotically flat Schwarzschild black holes. In fact, they are thermodynamically stable, and the general principles introduced by Gubser and Mitra \cite{kn:gubsermit} would then indicate that they should also be dynamically stable. \emph{Thus, the combination of holography with Figure 1 is in fact asking for some very unusual form of black hole instability to occur along the line BC.} This is why it is difficult to account for BC holographically.

The situation begins to become clearer if we consider asymptotically AdS black holes with \emph{flat} event horizons having the topology of [say] a torus; this is in any case a very natural move from the holographic point of view, since we are really interested in field theories defined on flat spacetimes, not spatially spherical ones. Such black holes do not exist in the asymptotically flat case, but, significantly, they do exist in the asymptotically AdS environment \cite{kn:lemmo}. Like spherical AdS black holes, these ``flat" black holes undergo a first-order Hawking-Page phase transition when their temperature is lowered \cite{kn:surya}. Below this temperature, the black hole is replaced by another geometry [a \emph{Horowitz-Myers soliton} \cite{kn:horomy}]. But in the \emph{flat} case this happens at the same temperature at all values of the chemical potential. In fact the black hole must satisfy
\begin{equation}\label{ALEPH}
\m{T\; \geq \; {1 \over 2\pi KL},}
\end{equation}
where T is the Hawking temperature, 2$\pi$K is the periodicity of the torus [assumed cubic], and L is the asymptotic AdS curvature scale. This means that the dual field theory plasma can never be arbitrarily cold, at any value of the chemical potential; which is reasonable. Of course, other forms of instability may intervene before the temperature in (\ref{ALEPH}) is reached, and, as we shall see, this does happen in some circumstances; also, the actual description of the transition involves more complicated objects such as black holes with scalar ``hair"; but this merely reinforces the point ---$\,$ stable flat AdS Reissner-Nordstr$\ddot{\m{o}}$m black holes cannot be arbitrarily cold. This resolves the apparent conflict with the third law of thermodynamics: zero temperature cannot in practice be attained or even neared by the relevant black holes, so the question does not arise\footnote{For an alternative approach to the problem, not motivated by holography, see \cite{kn:dhoker}.}.

Thus we have an improvement over the situation in the case of black holes with spherical event horizons: for example, it is no longer possible to cause the hadronic state to make a transition to a plasma, at arbitrarily low temperatures, simply by increasing the chemical potential. However, we still see no sign of the triple point; the theory seems to indicate that the field theory makes a transition to a confined state, as the temperature is lowered, at all values of the chemical potential. At the other extreme, we see no sign of the critical point. Figure 1 evidently demands that some other forms of instability must arise and supplant this one at very small, and also at very large, values of $\mu$.

One can try to make progress by embedding the entire discussion more completely in string theory. Many new forms of black hole instability then arise: see for example \cite{kn:gubsermit}\cite{kn:yamada1}\cite{kn:yamada2}. As mentioned in the previous section, it is reasonable to hope that the rapidly growing understanding of holographic superconductivity will allow us to use these or related forms of black hole instability to account for the region of Figure 1 around the critical point. We therefore confine our attention to the triple point, B, and the associated phase line BC.

In this work, we investigate the possibility, discussed briefly in \cite{kn:AdSRN}, that the black hole instability defining the line BC in Figure 1 is due to the brane pair-production instability discovered by Seiberg and Witten \cite{kn:seiberg}\cite{kn:wittenyau}\cite{kn:maoz}\cite{kn:porrati}\cite{kn:conspiracy}. This is a strong candidate for the effect defining BC for the following reason: this form of instability can afflict black holes \emph{even at very high temperatures}, provided that the electric charge is sufficiently large [but still sub-extremal]. As we discussed earlier, this reflects the actual behaviour of the QGP, to the extent that it is currently understood. Since the triple point does undoubtedly lie at very high values of the chemical potential, Seiberg-Witten instability certainly cannot be ignored in this region of the phase diagram.

In \cite{kn:AdSRN}, once again, the key point is that the relevant black holes have \emph{flat} event horizons. The spacetime at infinity is therefore exactly flat [and so its Euclidean version does not have positive scalar curvature, as it does in the case of spherical AdS black holes], and as a result these black holes are perilously close to being unstable to the brane pair-production effect discussed by Seiberg and Witten. It turns out that when the charge is increased to a value near, but still below, the extremal value, AdS Reissner-Nordstr$\ddot{\m{o}}$m black holes with flat event horizons do cease to be stable in this manner, and must make a transition to some other state, which is \emph{not} a Horowitz-Myers soliton. Requiring stability against the Seiberg-Witten effect excludes a sector of the black hole parameter space corresponding to the region under a \emph{diagonal} line in the phase diagram of the dual theory, given by the locus
\begin{equation}\label{BETH}
\m{T\;=\;{2\gamma \over \sqrt{3\pi }}\,\mu},
\end{equation}
where $\gamma$ is a positive constant parameter; this relation will be derived in section 3, below. \emph{We see at once that Seiberg-Witten instability can indeed affect extremely hot black holes}, provided that $\mu$ is large. This is exactly what we need.

\begin{figure}[!h]
\centering
\includegraphics[width=0.9\textwidth]{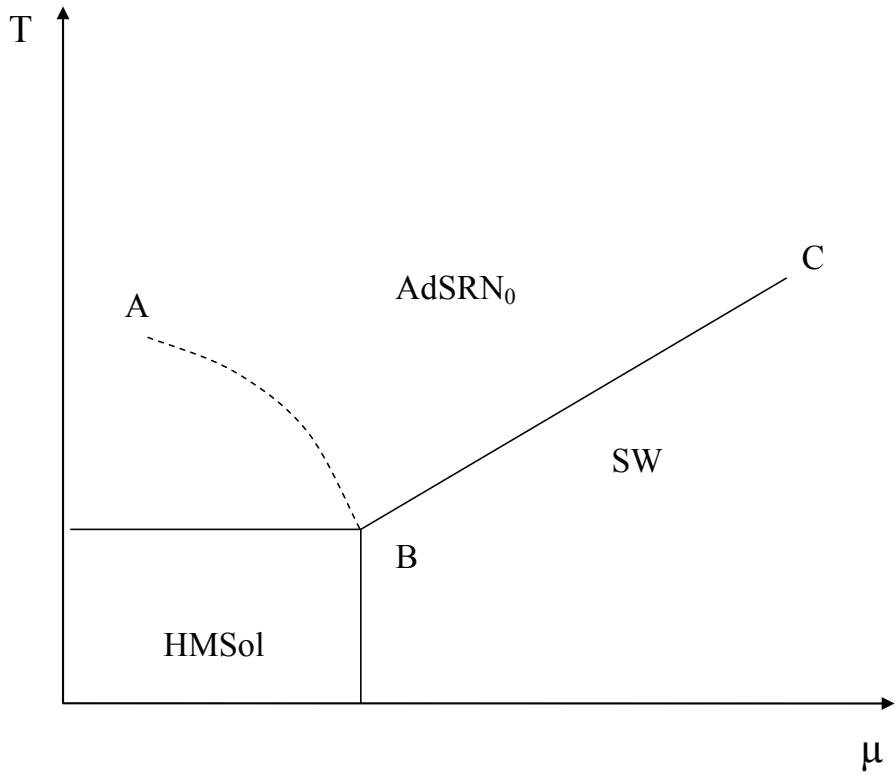}
\caption{Holographic Phase Diagram. AdSRN$_0$ = [stable] black hole phase, HMSol = Horowitz-Myers Soliton phase, SW = Seiberg-Witten phase. Dotted line: see Figure 3.}
\end{figure}

Let us for the moment leave aside other possible forms of black hole instability, and consider only two: the Hawking-Page transition we discussed earlier, and this Seiberg-Witten instability. As the temperature is lowered, the two kinds of instability compete: if one sets in first, it changes the geometry and renders the other irrelevant. The excluded parameter values correspond to regions of the phase diagram lying below two straight lines, one horizontal [inequality (\ref{ALEPH})] and one diagonal [equation (\ref{BETH})]. The question as to which effect dominates is determined by deciding which of the two lines is above the other. The horizontal line dominates at low chemical potentials, the diagonal line at large values of $\mu$. We thus obtain an [of course very rudimentary] holographic version of the phase plane sketched in Figure 1; see Figure 2. [In this figure, by the ``SW phase" we just mean the phase which cannot be described by AdS Reissner-Nordstr$\ddot{\m{o}}$m black holes, because of the Seiberg-Witten instability; its precise nature is not our concern here.]

Figure 2 certainly represents progress: at least now we have three phases and an obvious candidate for the triple point, and we have a concrete mechanism for explaining how very hot, highly charged AdS black holes can be unstable. Furthermore, by going beyond the simplest possible background ---$\,$ pure AdS Reissner-Nordstr$\ddot{\m{o}}$m geometry, albeit with an unusual boundary structure ---$\,$ one can hope to improve on this picture; one might try, for example, to apply the ideas of \cite{kn:denef} to the present situation. The hope is that the sketch in Figure 2 will then become more realistic, with curved instead of straight lines, and so on. [In fact, one now knows, as mentioned above, that the deconfinement/confinement transition at low values of $\mu$ is not really described by a simple Hawking-Page transition, but rather requires the inclusion of black holes with scalar ``hair"; this will have a strong effect on the horizontal line in Figure 2. This line plays no further role in the present work.]

Unfortunately, the critical point lies in a region [the crossover in Figure 1] that is not yet properly understood holographically; its location is for us an input from phenomenology and [near-future] experiment. Holographically, the critical point must correspond to values of the mass and charge of the AdS Reissner-Nordstr$\ddot{\m{o}}$m black hole which are very ``special", in some as yet unknown way. Despite the fact that we do not understand the special character of what we shall call the ``\emph{critical black hole}", we can still make use of it, in the following way. By modifying the charge/mass ratio Q/M of the dual black hole, we can define trajectories in the holographic phase plane, beginning for example at the critical point. We shall see that increasing Q/M gives rise to a trajectory which bends down and to the right, as indicated by the dotted line in Figure 2. Eventually this trajectory terminates when Seiberg-Witten instability sets in.

Now the terminal point of the trajectory which begins at the critical point has a special status: it corresponds to the coldest black hole that can be obtained in this way, starting at a black hole with a definite QGP interpretation. \emph{We therefore propose to identify it with the triple point.} The idea is then to use what is known about the position of the critical point to locate the triple point; we wish to see whether the resulting prediction is reasonable. The assumption being tested here is that Seiberg-Witten instability describes the breakdown of the QGP state at extreme values of T and $\mu$, along the line BC in Figures 1 and 2. Other forms of instability, such as those discussed in \cite{kn:gubsermit}\cite{kn:yamada1}\cite{kn:yamada2}\cite{kn:horror}\cite{kn:horrorrob}\cite{kn:rocha}\cite{kn:kachru}\cite{kn:aprile}\cite{kn:pani}, are relevant in the region around the critical point. \emph{In short, the more familiar forms of black hole instability should account for the region of relatively small $\mu$, leaving the Seiberg-Witten effect to account for extremely large values, along BC}.

Because of the presence of the unknown parameter $\gamma$ in equation (\ref{BETH}), it turns out that we can only use this idea to constrain the temperature of the triple point to the extent of putting a lower bound on it. However, we shall see that this lower bound actually almost coincides with a theoretical \emph{upper} bound obtained by Shuryak in \cite{kn:shuryak}. Thus we can fix the temperature, and, with some more work, $\gamma$. This in turn allows us to estimate the value of the chemical potential at the triple point, thus specifying the latter's position in the phase diagram.

To prepare for that computation, we need a brief review of the way in which the black hole parameters are related to the physical parameters of the quark plasma.

\addtocounter{section}{1}
\section* {\large{\textsf{3. Black Hole Parameters and their Duals}}}
The black hole metric in question is the AdS Reissner-Nordstr$\ddot{\m{o}}$m metric with a \emph{flat} event horizon \cite{kn:lemmo}:
\begin{equation}\label{D}
\m{g(AdSRN_0) = - \Bigg[{r^2\over L^2}\;-\;{2M\over
3\pi^2K^3r^2}\;+\;{Q^2\over 48\pi^5 K^6 r^4}\Bigg]dt^2\;+\;{dr^2\over {r^2\over L^2}\;-\;{2M\over
3\pi^2K^3r^2}\;+\;{Q^2\over 48\pi^5 K^6 r^4}} \;+\; r^2d\Omega_0^2}.
\end{equation}
Here we are taking the bulk to be a five-dimensional, asymptotically locally AdS spacetime with asymptotic curvature $- 1$/L$^2$; M and Q are the ADM mass and charge\footnote{For simplicity, all charges are taken to be positive in this work.}, and the event horizon is a perfectly flat cubic torus parametrised by angles with periodicity 2$\pi$K and with metric d$\Omega_0^2$. The field theory is then defined on a flat spacetime with formal periodic boundary conditions. We can think of M and [particularly] Q as quantities which we can manipulate, to some extent, by dropping elementary particles into the black hole. On the other hand, L and K cannot be modified by any local physical process.

The value of the radial coordinate at the event horizon is obtained by finding the larger of the positive real roots of the equation
\begin{equation}\label{G}
\m{{r_{eh}^6\over L^2}\;-\;{2Mr_{eh}^2\over
3\pi^2K^3}\;+\;{Q^2\over 48\pi^5 K^6} \;=\;0.}
\end{equation}
The quantity r$_{\m{eh}}$ can be replaced by the black hole entropy S, to which it is related in the usual way \cite{kn:lemos2}, that is, by computing one quarter of the [three-dimensional] area of the event horizon:
\begin{equation}\label{H}
\m{S \;=\;2\pi^3K^3r_{eh}^3.}
\end{equation}
Eliminating r$_{\m{eh}}$ between (\ref{G}) and (\ref{H}), we obtain
\begin{equation}\label{I}
\m{\pi Q^2L^2 \;=\;2^{1/3} \times 16 \pi^2 M L^2 K S^{2/3} \;-\; 12S^2.}
\end{equation}
This equation allows us to fix all of the geometric and ADM parameters but one, and then to regard S as a function of the remaining parameter or vice versa. For example, one could fix all of the parameters but the mass; S and M are then increasing functions of each other [in the domain of these parameters which corresponds to the correct choice of the root of (\ref{G}), the one defining r$_{\m{eh}}$]. Alternatively, one can fix all of the parameters but the charge; S and Q are then \emph{decreasing} functions of each other.

The temperature of the black hole, corresponding to its Hawking radiation, is computed from the circumference of the circle defined by Euclidean ``time"; the result \cite{kn:AdSRN} is
\begin{equation}\label{JJ}
\m{T_{}\;=\;{1\over  2^{1/3}\pi}\Bigg[{S^{1/3}\over \pi KL^2}\;-\;{Q^2 \over 24K S^{5/3}}\Bigg].}
\end{equation}
If we fix all parameters but the mass, this is an increasing function of S and hence of M: increasing the mass of these black holes \emph{increases} their temperature, contrary to the more familiar behaviour of [asymptotically flat] Schwarzschild black holes. These black holes have positive specific heat; if they are not unstable in some other way, they can attain thermal equilibrium with their own Hawking radiation: they are eternal \cite{kn:juan}\footnote{Our discussion in Section 2, to the effect that one would \emph{expect}  highly charged spherical black holes to be very stable, therefore applies here also.}. We shall always assume initially that all of the black holes with which we deal are static in this sense, so that, in particular, the temperature is not time dependent.

If we fix all parameters but the charge, the temperature is again an increasing function of S, and so it is a decreasing function of Q: increasing the charge \emph{causes the temperature to drop}. In this case we have
\begin{equation}\label{J}
\m{T\;=\;{3S^{1/3}\over 2^{4/3}\pi^2 KL^2}\;-\;{2M\over 3S}.}
\end{equation}
The first term dominates at large values of S, and gives the expected relation between T and S; but at small values of S [that is, large charges], the second term becomes important and forces the temperature to vanish at a non-zero value of S. The black hole can apparently have an arbitrarily small temperature, though its entropy is bounded away from zero: this is the phenomenon discussed in Section 1.

In \cite{kn:AdSRN} we showed that arbitrarily low temperatures do \emph{not} occur, because asymptotically AdS black holes with flat horizons, unlike their spherical counterparts, become unstable to the Seiberg-Witten effect [discussed in Section 2 above] when their charges reach a certain ``\emph{near-extremal}" value, Q$_{\m{NE}}$, which is related to the extremal charge Q$_{\m{E}}$ by
\begin{equation}\label{CRAPA}
\m{Q_{NE}/Q_E\;=\;{3^{3/4}\over 2^{5/4}}\;\approx \;0.958415.}
\end{equation}
It follows that there is a minimal possible [``near-extremal"] temperature [for fixed values of all parameters but the charge]
T$_{\m{NE}}\,>\,0$, defined by Q$_{\m{NE}}$; from \cite{kn:AdSRN} we have
\begin{equation}\label{T}
\m{T_{NE}\;=\;{1\over 2\times 3^{1/4}\times \pi^{3/2}}\,\Bigg[{M\over K^3L^6}\Bigg]^{1/4}}.
\end{equation}
\emph{The black hole becomes unstable if one tries to go below this temperature by increasing the charge.}

The energy density of the field theory is proportional to the mass per unit horizon area of the dual black hole; that is, it is proportional to M/4S. The entropy of the black hole has a minimal value S$_{\m{NE}}$, corresponding to the ``near-extremal" temperature. It is slightly larger than the entropy of the extremal hole, and is given by
\begin{equation}\label{ZEBRA}
\m{S_{NE}\;=\;{2\over 3^{3/4}}\times \Big[\pi^2 M L^2 K\Big]^{3/4}.}
\end{equation}
If we fix all of the parameters except the charge, then, M/4S has a maximal value given by M/4S$_{\m{NE}}$. The ratio of the density at any value of the charge to its maximal density is therefore
\begin{equation}\label{CRAPPY}
\m{\rho/\rho_{max}\;=\;S_{NE}/S}.
\end{equation}
It is interesting to construct a fundamental dimensionless quantity $\sigma^{*}$, defined by
\begin{equation}\label{CRAPB}
\m{\sigma^*\;=\;{S_{NE} T_{NE}\over M}}.
\end{equation}
This is the obvious quantity to consider from the point of view of the fundamental thermodynamic relation. Intuitively, in the dual picture, it measures the minimal possible temperature of the quark plasma in terms of the corresponding ``near-extremal" energy density. Classically, it is, as we have seen, equal to zero; in string theory, this quantity can be computed and is non-zero \cite{kn:AdSRN}: it is given simply by
\begin{equation}\label{CRAPC}
\m{\sigma^*\;=\;1/3}.
\end{equation}
The fact that this number is not particularly small gives a general indication that the lowest possible temperature of the plasma [for fixed parameter values apart from the charge] may not be unreasonably low relative to the scale defined by the energy density.

The electromagnetic potential in this spacetime is given by [see \cite{kn:AdSRN} for the details]
\begin{equation}\label{L}
\m{A\;=\;{Q \over 16\pi^3K^3}\Bigg[{1\over r^2}\;-\;{1\over r_{eh}^2}\Bigg]\,dt.}
\end{equation}
The chemical potential $\mu$ of the field theory is proportional to the magnitude of the asymptotic value of the coefficient in this expression. As explained in  \cite{kn:myers}, the constant of proportionality must have units of inverse length. Let us take it to be 1/$\gamma$L, where L is the asymptotic curvature scale as usual, and $\gamma$ is dimensionless. [This quantity plays a crucial role in the sequel; we shall shortly return to the question of fixing its value.] Then
\begin{equation}\label{M}
\m{\mu\;=\;{Q \over 16\pi^3K^3\gamma L r_{eh}^2}\;=\;{2^{2/3}Q \over 16\pi K \gamma L S^{2/3}}.}
\end{equation}
By inserting the ``near-extremal" values of Q and S [see equation (\ref{ZEBRA})] into this formula, we obtain the maximal possible value of the chemical potential:
\begin{equation}\label{MAMA}
\m{\mu_{NE}\;=\;{3^{1/4}\over 4 \pi \gamma}\,\Bigg[{M\over K^3L^6}\Bigg]^{1/4}}.
\end{equation}
\emph{The black hole becomes unstable in the Seiberg-Witten sense if this value is exceeded.}

Eliminating Q, just as we did in the case of the temperature, we find
\begin{equation}\label{MONSTER}
\m{\mu^2\;=\;{3\times 2^{2/3}\,S^{1/3}\over 16\pi \gamma^2 KL^2}\,\Bigg[{2M\over 3S}\;-\;{S^{1/3}\over 2^{4/3}\pi^2 KL^2}\Bigg].}
\end{equation}
This can be compared with the analogous formula (\ref{J}) for the temperature. If we change the charge to mass ratio Q/M by increasing Q while fixing M, then a given point in the ($\mu$, T) plane will move along a trajectory given parametrically, with S as the parameter, by (\ref{MONSTER}) and (\ref{J}). Two trajectories of this kind, with arbitrarily prescribed coefficients, are pictured in Figure 3; the upper curve is obtained from the lower by increasing M. The dotted line in Figure 2 is based on these graphs.
\begin{figure}[!h]
\centering
\includegraphics[width=1.0\textwidth]{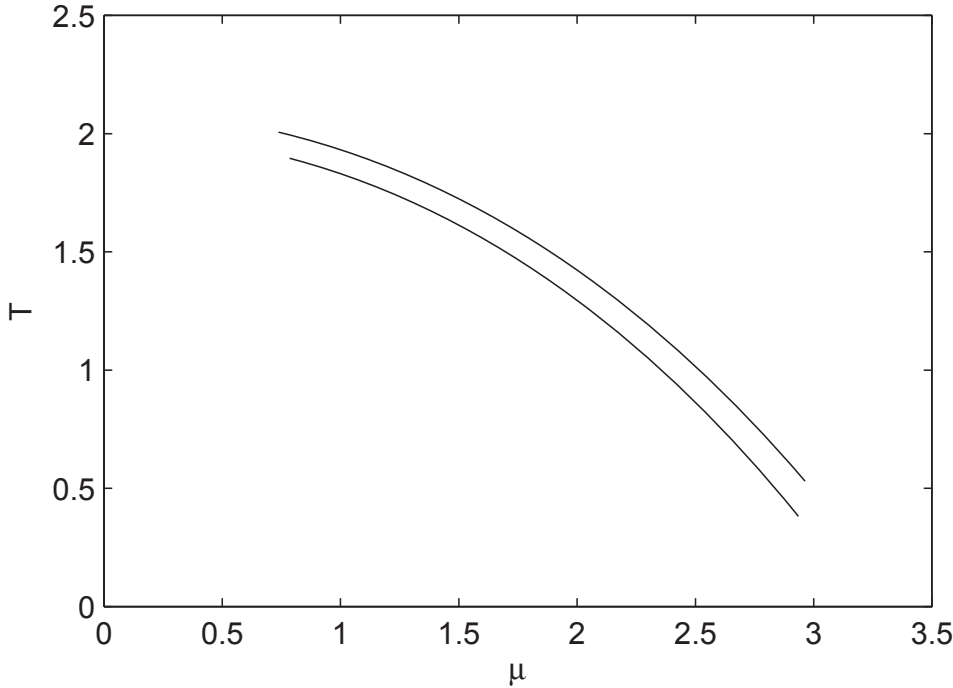}
\caption{Trajectories of Black Holes in the Phase Diagram With Two Values of the Mass.}
\end{figure}

In principle, but unfortunately not yet in practice, the value of $\gamma$ can be derived from string-theoretic considerations \cite{kn:kob}\cite{kn:myers}; in the latter reference, and, following it, in \cite{kn:AdSRN}, $\gamma$ is taken to be equal to $\pi$, but this was for the sake of simplicity: ``$\mu$" in those references should be read as ``$\mu\gamma/\pi$". Clearly, $\gamma$ plays a vital role in linking the two sides of the AdS/CFT duality, as we are applying it here, and it is essential to be able to constrain or determine it. For example, it plays a basic role in string-theoretic corrections to the celebrated ``\emph{KSS bound}" \cite{kn:son} on the viscosity/entropy density
ratio $\eta$/s.

To see how this works, let us, following \cite{kn:myers}, define a dimensionless temperature-normalized chemical potential,
\begin{equation}\label{P}
\m{\bar{\mu}\;=\;\mu/T.}
\end{equation}
This parameter\footnote{As above, ``$\bar{\mu}$" in \cite{kn:myers}\cite{kn:AdSRN} should be read as ``$\bar{\mu}\gamma/\pi$".} is of interest because it is \emph{directly} related to observational data on the matter/antimatter asymmetries observed, for example, at the highest energy collisions in the RHIC experiment \cite{kn:phobos}.  The KSS bound ---$\,$ given by 1/4$\pi$ ---$\,$ holds for a wide variety of systems, provided that one uses the standard two-derivative action for bulk gravity, but it can fail if higher-derivative terms, required by string theory, are taken into account \cite{kn:ge}. Furthermore, this failure is in some cases enhanced by the inclusion of a chemical potential \cite{kn:myers}\cite{kn:cremonini}, which enters precisely in the form of $\bar{\mu}$: according to Myers et al., the KSS ratio is given, in a four-derivative theory, by
\begin{equation}\label{A}
\m{{\eta\over s}\;=\;{1\over 4\pi}\,\Bigg[1\;-\;8c_1 \;+\;G(\gamma\bar{\mu})\,\times\,\Big(c_1\;+\;6c_2\Big)\Bigg]},
\end{equation}
where c$_1$ is the dimensionless coupling of the squared curvature term $\m{R_{abcd}R^{abcd}}$, c$_2$ is the dimensionless coupling of a term of the form $\m{R_{abcd}F^{ab}F^{cd}}$ [here F$\m{_{ab}}$ is the field strength tensor], and G($\bar{\mu}\gamma$) is the function
\begin{equation}\label{ALBANIA}
\m{G(\bar{\mu}\gamma)\;=\;{16\bar{\mu}^2\gamma^2/\pi^2\over 3\Big(1\;+\;\sqrt{1\;+\;2\bar{\mu}^2\gamma^2/3\pi^2}\Big)^2}}.
\end{equation}
Myers et al. \cite{kn:myers} give concrete examples in which both corrections in (\ref{A}) are negative. As G($\bar{\mu}\gamma$) is increasing, large values of $\bar{\mu}\gamma$ lead to more severe violations of the KSS bound. The value of $\bar{\mu}$ observed at the highest energy collisions in the RHIC experiment is quite small, about 0.16, but its value at the critical point is much higher, and at the triple point higher still: values around 15 or even a little beyond would not be considered outlandish at the triple point.

Now in fact the observed value of $\eta$/s is very close to 1/4$\pi$; certainly it is not far below it \cite{kn:teeny}. But as future experiments probe more deeply into the phase diagram, the corresponding large variations in $\bar{\mu}$ could conceivably lead to observable violations of the KSS bound. \emph{Whether this will happen depends in part on the value of $\gamma$}.
If $\gamma$ is around [say] 5, then, as $\bar{\mu}$ varies from 0.16 to 15, G($\bar{\mu}\gamma$) varies from 0.08464 to 7.22020; this latter figure represents a very substantial enhancement of the KSS bound violation, and this could perhaps be observable if the constants c$_1$ and c$_2$ are not too small. But if $\gamma$ is only [say] 0.10, then G($\bar{\mu}\gamma$) varies, over the same range of $\bar{\mu}$ values, only from 3.45840$\,\times\,10^{- 5}$ to 0.28285. Thus in this second case we would predict that violations of the KSS bound will \emph{not} be observed in the quark-gluon plasma even near to the triple point. The importance of determining $\gamma$ is clear.

In fact, theoretical and lattice computations concerning the critical point, combined with the requirement of stability against the Seiberg-Witten effect, impose a strong constraint on $\gamma$. Combining equations (\ref{T}) and (\ref{MAMA}), we see that Seiberg-Witten instability sets in if a point in the phase diagram moves below the line given by equation (\ref{BETH}). It follows that $\bar{\mu}$ and $\gamma$ have to be related by
\begin{equation}\label{BETHANY}
\m{\gamma\;\leq\;{\sqrt{3\pi}\over 2\bar{\mu}}}.
\end{equation}
A variety of theoretical, lattice, and experimental results [see for example \cite{kn:mohanty}] suggest that the QCD critical point lies at a temperature of about 150 MeV, and at a baryonic chemical potential of about 350-450 MeV. That is, the value of $\bar{\mu}$ is already about 2 to 3 there. Accepting this latter value for definiteness, we see that $\gamma$ can be no larger than about 0.51. Since we do not expect the triple point [where, in our picture, Seiberg-Witten instability actually occurs] to be very close to the critical point, and since it lies at lower values of the temperature but at higher values of the chemical potential [so that $\bar{\mu}$ is larger there], the actual value must be considerably smaller. As we shall see later, the temperature of the triple point is not very sensitive to $\gamma$ once we know that it is this small, so even this very rough estimate is useful: furthermore, already it seems most unlikely that investigations of the quark-gluon plasma at high chemical potentials will reveal any violations of the KSS bound.

To return to the discussion of Figure 3: we can summarize as follows. Since S is a decreasing function of Q,  $\mu$ is an increasing function of Q when the other parameters are fixed [and so is $\bar{\mu}$]; similarly, $\mu$ and $\bar{\mu}$ are decreasing functions of M when Q is fixed [see equation (\ref{M})].
We see, then, that increasing the mass of the black hole tends to move a point in the phase diagram, corresponding to a given state of the quark plasma, upwards and to the left; this can be seen in Figure 3, where the upper curve is obtained from the lower by increasing M. Increasing the charge moves the point downwards and to the right: this too can be seen in Figure 3. In short, dropping particles with a large charge/mass ratio into a black hole of this kind will give rise to a trajectory in the phase diagram like the ones shown in Figure 3 ---$\,$ or, indeed, like the dotted line in Figure 2. Let us try to use this to estimate the location of the triple point.

\addtocounter{section}{1}
\section* {\large{\textsf{4. From the Critical Point to the Triple Point}}}
We have seen that, if the ratio Q/M is increased by fixing M and increasing Q, the point representing the black hole will descend in the phase diagram along a path like the ones shown in Figure 3; its parametric form is given by equations (\ref{MONSTER}) and (\ref{J}).

In fact, if we try to manipulate the ADM parameters of a black hole in a physical way, by dropping elementary particles into it, we will be changing the hole's mass and charge in almost precisely this way. For charged elementary particles have charges which [in Planck units] vastly exceed their masses\footnote{It should be noticed in this connection that, in string theory, the relative weakness of gravitation is probably built into the theory; see \cite{kn:vafa}.}, so in effect Q increases while M remains constant. This process can continue until extremality is attained, unless the black hole becomes unstable first ---$\,$ which, as we have seen, will indeed happen when the event horizon is flat.

The critical point A in Figure 1 corresponds to specific values of the temperature and chemical potential. In the dual picture, this means that there is a particular black hole ---$\,$ the critical black hole ---$\,$ with very ``special" values of the mass and charge. By increasing the charge on the critical hole, one generates a trajectory which intersects the line (\ref{BETH}) at a certain point. Starting with a more massive black hole, one finds numerically [adjusting M in (\ref{MONSTER}) and (\ref{J})] that the trajectory intersects that line at a higher temperature; this is clear also from Figure 3. Less massive black holes yield a lower temperature at the intersection, but these holes are dual to points \emph{below} the phase line starting at A in Figure 1, so they do not correspond to the QGP. We see that the point determined in this manner by beginning at the critical point represents \emph{the coldest black hole which can be obtained by cooling a black hole with a QGP dual} in a physically meaningful way. Since the triple point is the coldest possible state of the QGP, it seems very natural to identify the triple point with the intersection of the critical point's trajectory with the line of Seiberg-Witten instability, and this is what we are proposing here.

Our objective is to find, or at least to constrain, the temperature and chemical potential at the point where the trajectory beginning at the critical black hole intersects the line given by equation (\ref{BETH}). This is a complicated algebraic problem; we proceed as follows.

Equation (\ref{J}) can be re-written in the following way:
\begin{equation}\label{K}
\m{K\;=\;{9\,S^{4/3} \over 2^{1/3}\times 4\pi^2 ML^2(1\;+\;3TS/2M)};}
\end{equation}
this serves as a useful reminder that the spatial periodicity parameter K depends on the physical parameters of the black hole, in particular its temperature. Note that, in the Euclidean version of the spacetime, the conformal boundary is a four-dimensional torus with periodicity fixed by TL in the ``time" direction, and by K in the other three; thus the black hole temperature has a geometric interpretation in terms of the shape of the torus at infinity.

We can relate K to physical parameters in another way: solving equation (\ref{MONSTER}), we obtain after extensive simplifications
\begin{equation}\label{N}
\m{K\;=\;{2^{2/3}\times 3\,S^{4/3}/4\pi^2\,ML^2 \over 1\;+\;\sqrt{1\;-\;(12\gamma^2 \mu^2S^2/\pi M^2)}};}
\end{equation}
combining this with equation (\ref{K}) we will be able to eliminate K from our discussion entirely\footnote{One can show that $\m{12\gamma^2 \mu^2 S^2/\pi M^2}$, regarded as a function of S, has a maximum possible value of 1, so that there are no difficulties with the square root.}.

It is useful to use a dimensionless parameter $\sigma$, defined by
\begin{equation}\label{O}
\m{\sigma \; = \; TS/M.}
\end{equation}
This is of course motivated by equation (\ref{CRAPB}); as in the discussion of that equation, $\sigma$ is a measure of the temperature relative to the reciprocal of the entropy of the black hole, or to the energy density of the field theory. Since T is an increasing function of S, so is $\sigma$, and so $\sigma$ is a decreasing function of the black hole charge. The requirement of stability against the Seiberg-Witten effect means that $\sigma^*$ = 1/3 is in fact the smallest possible value of $\sigma$.

Equating the two expressions for K in (\ref{K}) and (\ref{N}), and using the definitions of $\sigma$ and $\bar{\mu}$ [equation (\ref{P})], we obtain at length
\begin{equation}\label{Q}
\m{9\Big(1\;+\;{12\gamma^2\bar{\mu}^2\over \pi}\Big)\sigma^2\;-\;6\sigma\;-\;8\;=\;0.}
\end{equation}
Thus in fact $\sigma$ can be computed, as the positive root of this quadratic, once $\bar{\mu}$ and $\gamma$ are specified; for example, one can compute $\m{\sigma_{critical}}$, the value at the critical point, by inserting a value for $\gamma$ and the number [$\approx$ 3] we have been using for $\bar{\mu}$ at the critical point.

One sees from (\ref{Q}) that $\sigma$ is a decreasing function of $\gamma$: in fact, the inequality (\ref{BETHANY}) in the preceding section can be derived from this equation and from the fact that $\sigma$ is bounded below by 1/3. In addition, however, equation (\ref{Q}) means that there is an \emph{upper} bound on $\sigma$: when $\gamma$ tends to zero, we find that $\sigma$ approaches 4/3. Thus we have
\begin{equation}\label{QUEST}
1/3\;\leq \;\sigma \; < \;4/3.
\end{equation}

Replacing S by $\sigma$ in equation (\ref{J}), and then solving for T, we obtain
\begin{equation}\label{S}
\m{T\;=\;{3^{3/4}\sigma^{1/4}M^{1/4}\over 2\pi^{3/2}(KL^2)^{3/4}[1\;+\;2/(3\sigma)]^{3/4}}}\;.
\end{equation}
Now the minimum temperature below which the black hole becomes unstable is given by the temperature at the ``near-extremal" value of the charge: see equation (\ref{T}) above.
According to our hypothesis, this ``near-extremal" temperature in the case of our distinguished trajectory in the phase diagram is just the temperature at the triple point. Re-labelling it accordingly and combining (\ref{T}) with (\ref{S}), we have, since this trajectory begins at the critical point,
\begin{equation}\label{U}
\m{T_{triple}\;=\;{1\over 3\sigma_{critical}}\Bigg({2\over 3}\;+\;\sigma_{critical} \Bigg)^{3/4}T_{critical}}.
\end{equation}

Next, equation (\ref{BETH}) allows us to compute the chemical potential at the triple point in terms of $\gamma$ and the temperature there:
\begin{equation}\label{UU}
\m{\mu_{triple}\;=\;{\sqrt{3\pi}\over 2\gamma}\,T_{triple}}.
\end{equation}

Finally, the energy density of the field theory is proportional to the mass per unit horizon area of the black hole; since the entropy is one quarter of the area, this means that the relevant quantity is M/4S, or T/4$\sigma$. Thus the energy densities at the critical and triple points obey
\begin{equation}\label{V}
\m{\rho_{triple}/\rho_{critical}\;=\;{\sigma_{critical} T_{triple} \over T_{critical} \sigma_{triple}}},
\end{equation}
where, as we have discussed, $\sigma_{\m{triple}}$ is 1/3. Thus in fact
\begin{equation}\label{W}
\m{\rho_{triple}/\rho_{critical}\;=\;\Bigg({2\over 3}\;+\;\sigma_{critical} \Bigg)^{3/4}}.
\end{equation}

As we have seen, a precise determination of $\sigma$ depends having a value for $\gamma$, which is not currently available from a first-principles calculation. However, we can make progress in two ways. First, the upper bound on $\sigma$ given in (\ref{QUEST}) means that \emph{we have a lower bound on the temperature of the triple point}: we have \begin{equation}\label{X}
\m{T_{triple}\;\geq\;{2^{3/4}\over 4}\,T_{critical}}.
\end{equation}
Using our standard estimate of the temperature of the critical point, namely 150 MeV, we have a bound which does not depend on $\gamma$: approximately,
\begin{equation}\label{Y}
\m{T_{triple}\;\geq\; 63 \;MeV}.
\end{equation}
Similarly, the upper bound on $\sigma$ yields an upper bound on the energy density corresponding to the triple point:
\begin{equation}\label{Z}
\m{\rho_{triple}/\rho_{critical}\;\leq\;(2)^{3/4}}.
\end{equation}
The energy density at the critical point is estimated \cite{kn:mohanty} to be not far below the maximal density attained in the RHIC experiment\footnote{The precise meaning of ``maximal energy density" in heavy ion physics is a somewhat subtle question, explained very clearly in \cite{kn:phobos}.}. For the sake of definiteness we shall therefore take  $\rho_{\m{triple}}$ to be around 1000 MeV/fm$^3$; other values can be accommodated in the obvious way\footnote{By way of comparison, note that 1 GeV/fm$^3$ = 1.78 $\times 10^{18}$ kg/m$^3$, while nuclear density is about 2.8 $\times 10^{17}$ kg/m$^3$.}. Using this estimate, we have, again approximately,
\begin{equation}\label{ZZ}
\m{\rho_{triple}\;\leq \;1680 \; MeV/fm^3}.
\end{equation}

The inequalities (\ref{Y})(\ref{ZZ}) have the virtue of being strong predictions, in the sense that they are independent of the value of $\gamma$; but they of course do not amount to fixing the location of the triple point in the phase diagram. We can work towards that goal by examining some numerical data. The table shows how the temperature, chemical potential, and energy density of the triple point vary as $\gamma$ is reduced, $\sigma_{\m{critical}}$ being computed from the basic equation (\ref{Q}). [Units are either MeV for temperature and chemical potential, or MeV/fm$^3$ for density; the first row is merely a reminder of the assumed  values at the critical point itself ---$\,$ that is, at these values the critical and triple points would coincide.]

\begin{center}
\begin{tabular}{|c|c|c|c|c|}
  \hline
$\gamma$ & $\sigma_{\m{critical}}$  &  T$_{\m{triple}}$ & $\mu_{\m{triple}}$ & $\rho_{\m{triple}}$ \\
\hline
0.51 &    1/3      &  150         &  450        &   1000.00 \\
0.45 &    0.37862  &  136.519   &  465.677  &   1033.78 \\
0.40 &    0.42461  &  125.728   &  482.480  &   1067.70 \\
0.35 &    0.48189  &  115.116   &  504.863  &   1109.47 \\
0.30 &    0.55444  &  104.756   &  535.998  &   1161.63 \\
0.25 &    0.64764  &  94.767    &  581.866  &   1227.51 \\
0.20 &    0.76710  &  85.344   &  655.010  &   1310.88 \\
0.15 &    0.92044  &  76.812    &  786.039  &   1414.02 \\
0.10 &    1.09836  &  69.709    &  1070.025 &   1531.31 \\
0.08 &    1.16946  &  67.439    &  1293.981 &   1577.34 \\
0.06 &    1.23416  &  65.585    & 1677.873  &   1618.85 \\
0.04 &    1.28671  &  64.206    & 2463.904 &    1652.30 \\
0.02 &    1.32126  &  63.355    &  4862.481 &   1674.17 \\
  \hline

\end{tabular}
\end{center}

Now Shuryak \cite{kn:shuryak} has given an ingenious argument, based on ``universal" aspects of cold fermionic atoms and quark matter, which supplies an \emph{upper} bound on the temperature of the triple point. This upper bound is around 70 MeV. A combination of our bounds with Shuryak's therefore gives us
\begin{equation}\label{ZZZZ}
\m{63\;MeV\;\leq \; T_{triple}\;\leq\; 70\;MeV;\;\;\;\;1530\;MeV/fm^3\;\leq \;\rho_{triple} \;\leq \;1680 \; MeV/fm^3}.
\end{equation}
The table allows us to translate Shuryak's bound to a rough upper bound on $\gamma$: this bound is around 0.10 to 0.15, with the lower end of this range being favoured. On the other hand, Shuryak's work also yields a rough estimate of the \emph{slope} of the phase line beginning at the triple point: it is about 0.1. Now according to equation (\ref{BETH}), the slope is determined by $\gamma$ via the equation
\begin{equation}\label{ZOSTER}
\m{{dT\over d\mu}\;=\;{2 \gamma\over \sqrt{3 \pi}}},
\end{equation}
which leads to values in the range $\approx \, 0.065$ --$\,$ 0.098 as $\gamma$ varies between $\approx \, 0.10$ --$\,$ 0.15. This is in rough agreement with Shuryak's estimate, but now it favours the upper end of the range for $\gamma$. This suggests that the actual value of $\gamma$ does lie in this range. For definiteness, let us fix $\gamma$ at 0.10: this allows both a refinement of our estimated temperature and also an estimate of the chemical potential at the triple point:
\begin{equation}\label{ZZZ}
\m{T_{triple}\;\approx\; 70\;MeV;\;\;\;\;\mu_{triple}\;\approx \; 1100 \;MeV;\;\;\;\;\rho_{triple} \;\approx \;1500 \; MeV/fm^3}.
\end{equation}
This means that $\bar{\mu}$ is quite large near to the triple point, around 15.7. [Our arguments imply that this value can never be exceeded anywhere in the region of the phase diagram corresponding to the quark-gluon plasma.] On the other hand, our predicted value for $\gamma$ is so small that, despite this, we predict that violations of the KSS bound [see the discussion around equations (\ref{A}) and (\ref{ALBANIA}) above] will not be observed even if this region of the phase diagram is eventually scanned experimentally.

Having fixed the location of the triple point, we can now describe the ``holographic quark matter phase diagram" in the immediate vicinity of that point. The general shape is given by the part of Figure 2 near to the point B; this point has, according to our proposals, coordinates $\mu$ $\approx$ 1100 MeV, T $\approx$ 70 MeV. At this point there meet three phases: the confined state, represented by Horowitz-Myers solitons [or by a suitable generalization derived from the theory of holographic superconductors], a ``Seiberg-Witten" phase which must be described by more complex models [perhaps like the one used in \cite{kn:chen} to give a holographic theory of the colour-flavour-locked state], and the QGP phase described by pure AdS Reissner-Nordstr$\ddot{\m{o}}$m black holes. From B there extends upwards and to the right an approximately straight line, representing a phase transition which continues to arbitrarily high values of the temperature and baryonic chemical potential. The slope of this line is predicted to be around 0.07 near the triple point.

\addtocounter{section}{1}
\section* {\large{\textsf{5. Conclusion}}}
Since the work of Seiberg and Witten \cite{kn:seiberg} it has been known that asymptotically AdS spacetimes can be very delicate objects when embedded in string theory. When the [Euclidean version of] the spacetime has negative scalar curvature at infinity, Seiberg and Witten showed that the bulk system is subject to a kind of pair-production instability involving branes. If [as is the case for AdS black holes with locally spherical event horizons] the scalar curvature at infinity is positive, and a certain positive energy condition holds, there is no such effect. Thus, the boundary between stable and unstable systems of this kind runs somewhere through the space of asymptotically AdS spacetimes with \emph{zero} scalar curvature at infinity\footnote{The conformal geometry allows us to treat the scalar curvature as a \emph{constant} \cite{kn:schoenyam}, and this yields the discrete classification of the boundary geometry into manifolds of strictly negative, positive, and zero scalar curvature.}.

In the holographic approach to understanding quark matter, the quark-gluon plasma is represented by asymptotically AdS black holes with zero scalar curvature at conformal infinity ---$\,$ that is, precisely by the delicate case in the Seiberg-Witten analysis.
One can show \cite{kn:conspiracy} that asymptotically AdS  black holes with flat event horizons are stable if their electric charge is zero. But it turns out that the addition of charge to such black holes distorts their bulk geometry [and therefore the rates of growth of the volumes and areas of branes as they propagate towards infinity]; this pushes them closer to the brink of instability in the Seiberg-Witten sense. One finds \cite{kn:AdSRN} that they do in fact become unstable when the charge is near to but still below the extremal value. \emph{This instability can affect even very hot, highly charged AdS black holes}, which one otherwise expects to be stable. In the dual picture, this imposes a constraint on the possible values of the temperature and chemical potential of the field theory; it forbids the QGP to exist even at extremely high temperatures if the baryonic chemical potential is very large.

In the present work, we have postulated that this restriction corresponds precisely to the phase transition which affects the quark-gluon plasma along the line BC in Figure 1, where indeed the QGP makes a phase transition at extreme values of T and $\mu$. We have used this idea to try to locate the quark matter triple point, using data on the critical point generally believed to exist not far from the region of the quark matter phase diagram currently being probed experimentally. The approach is based on the simple observation that the triple point represents the lowest possible temperature for the QGP. Combining these methods with the very different approach proposed by Shuryak \cite{kn:shuryak}, our conclusion is that this temperature is around 70 MeV.

Since even the critical point is currently poorly understood, the reader is entitled to question the numerical results we have obtained. Our main objective, however, is to show that the holographic approach, combined with very simple assumptions and techniques, may permit data about relatively well-understood regions of the quark matter phase diagram to be extended to probe regions which are otherwise almost entirely inaccessible. In view of the simplicity of the approach, the results obtained seem very reasonable: in particular, the ``lowest possible" temperature of the quark plasma is not absurdly low.

Clearly, however, much remains to be done. In particular, assuming that the holographic superconductivity programme does allow us to account for the region of the phase diagram around and below the critical point, one would like to patch this together with our holographic description of the region near and beyond the triple point. This will require an understanding of the relations between the more familiar kinds of black hole instability [involving scalar ``hair"] with the Seiberg-Witten variety. 

Another important issue concerns the fact that, since the black holes used here have toral event horizons, the dual gauge theory is also defined on a [spatial] torus. The consequences of that are well known in studies of lattice gauge theory, where the corresponding effects are known as ``finite-volume effects". A method of estimating the size of these effects is given in \cite{kn:harvey}; the basic question of course is to determine how large the torus must be taken in order to reduce the finite-volume effects to some prescribed level. In our case this would mean that the spatial periodicity parameter K must be taken to be sufficiently large. Note that our final results are apparently independent of K, but this may be misleading: intuitively one expects that ``large branes" [in the terminology of Seiberg and Witten themselves \cite{kn:seiberg}] will be slow to nucleate. This may mean in practice that Seiberg-Witten instability is more like a kind of metastability than an instability. Similar observations have been made by Hartnoll et al. \cite{kn:silverymoon} in their discussion of the ``Fermi seasickness" which may possibly afflict string-theoretic models of field theories at non-zero chemical potential. The consequences of this will depend on the relative spatial and temporal scales in any given application; that is, one would have to investigate whether, in a given case, there is enough time for the barrier to pair-production to be overcome. One might be able to do this by invoking causality and studying the distance from the black hole event horizon to the region where the brane action becomes negative. We hope to return to this issue elsewhere.

\addtocounter{section}{1}
\section*{\large{\textsf{Acknowledgement}}}
The author is grateful to Prof Soon Wanmei for help with numerical work.

\end{document}